\documentclass[11pt]{article}
\usepackage{formatting}
\usepackage{graphicx}
\usepackage{subcaption}
\usepackage{xcolor}

\title{Large-scale AI-Ready Data for Anti-Cancer Drug Response Modeling}
\author[1]{Vincent Lavelle}
\author[1,*]{Yitan Zhu}
\author[1]{Kaitlyn Marlor}
\author[2]{Thomas Brettin}
\author[2, 3]{Rick Stevens}
\affil[1]{Division of Data Science and Learning, Argonne National Laboratory}
\affil[2]{Computing, Environment, and Life Sciences, Argonne National Laboratory}
\affil[3]{Department of Computer Science, University of Chicago}

\affil[ ]{{*Correspondence:}  yitan.zhu@anl.gov}

\date{}

\begin{document}

\maketitle

\begin{abstract}
Drug response prediction (DRP) models are an active area of research in pharmacogenomics, with growing potential to accelerate the identification of effective anticancer drugs. However, their predictive performance is often constrained by limited dataset scale and insufficient coverages of cancer and chemical spaces. In addition, inconsistent benchmarking practices hinder reliable comparison across models. Standardized frameworks, such as the Innovative Methodologies and New Data for Predictive Oncology Model Evaluation (IMPROVE) project, provide unified data schemas and evaluation protocols for consistent benchmarking, but improving model generalizability requires larger and more diverse training data.

In this work, we substantially expand the IMPROVE benchmark through large-scale integration of pharmacogenomic data, primarily from PharmacoDB, together with additional smaller data sources. The expanded resource includes millions of drug–response measurements, broader multi-omics coverage, and a major increase in chemical diversity, adding more than 50,000 compounds. All data were curated and standardized using consistent data preprocessing, molecular featurization, and dose–response modeling procedures to produce a unified, AI-ready dataset suitable for drug response modeling.

To evaluate the impact of the new dataset compared to the original IMPROVE benchmark dataset, we trained DRP models using the two datasets and assess their prediction performance using a common test set and several evaluation strategies, including drug-blind, cancer-blind, and disjoint data splits. While cancer-blind performance remained comparable to the original benchmark, models trained on the expanded dataset showed consistent improvements in drug-blind and disjoint settings, indicating enhanced generalization to previously unseen compounds. These results position the expanded dataset as a community resource that provides a richer foundation for developing DRP models intended to aid in the discovery of novel anticancer drugs.

\end{abstract}

\section {1 Introduction}

Drug response prediction (DRP) models have become a prominent application of artificial intelligence (AI) and machine learning (ML) in pharmacogenomics, driven by the availability of large-scale in vitro drug screening and molecular profiling data \cite{Partin2023}. By integrating cancer gene expression profiles with chemical representations of therapeutic compounds, DRP models aim to learn predictive relationships that can inform drug discovery and patient treatment design. While methodological advances have continued to improve model architectures, the availability and structure of high-quality data remain central to progress in DRP research. 

Most DRP models are trained using pharmacogenomic datasets generated from cancer cell line drug screening experiments, where molecularly profiled cell lines are exposed to compounds across multiple doses and treatment effects are measured using viability-based assays. Over the past decade, several large studies have produced extensive collections of such data, and integrative resources such as PharmacoDB have played a critical role in curating and harmonizing these datasets across studies \cite{Smirnov2017}. By unifying cell line and compound identifiers and exposing drug response and molecular profiling data through a common interface, these resources have greatly improved accessibility and enabled cross-study analyses. 

However, pharmacogenomic data made available through such repositories are not inherently suited for large-scale AI workflows. Drug information is often provided as raw or non-canonical chemical representations, molecular profiles are derived from heterogeneous assays with different units and gene identifiers, and drug response measurements reflect study-specific preprocessing and modeling choices. As a result, substantial additional processing is required to transform these data into consistent numerical representations that can be directly consumed by modern ML models. Moreover, the lack of standardized data and unified interfaces complicates reproducible model training and evaluation across studies. The IMPROVE project was developed to address these challenges by establishing a standardized workflow for DRP model development and evaluation \cite{Weil2024}. By enforcing consistent preprocessing conventions, model interfaces, and evaluation protocols, IMPROVE enables reproducible benchmarking and systematic assessment of generalization behavior under diverse evaluation settings \cite{Partin2026}. While this framework provides a rigorous foundation for model comparison, its current benchmark dataset does not yet capture the full scale and diversity of available pharmacogenomic data.

To address this limitation, we extend the standardized data foundation underlying the IMPROVE workflow by constructing an expanded, AI-ready pharmacogenomic dataset. Building on data curated in PharmacoDB, we harmonize chemical, molecular, and drug response information into a unified representation aligned with the conventions used by IMPROVE \cite{Weil2024}. This includes transforming compound information into standardized molecular features, aligning data across sources using consistent identifiers and normalization strategies, and computing drug response metrics from raw experimental measurements using a uniform dose-response fitting scheme. The resulting dataset preserves the reproducibility and structure required for standardized benchmarking while substantially increasing the scale and diversity of data available for training DRP models. By releasing this resource to the community, we aim to support the development and evaluation of DRP models that more effectively leverage large-scale data within a rigorous and reproducible AI framework.

\section{2  Materials and Methods}

\subsection{2.1 Data Sources and Retrieval}

We retrieved data from multiple PharmacoDB constituent studies spanning diverse cancer cell line panels, compound libraries, and experimental protocols, including data from the following sources: Cancer Cell Line Encyclopedia (CCLE) \cite{Barretina2012,Ghandi2019}, Cancer Therapeutics Response Portal (CTRPv2) \cite{Seashore-Ludlow2015,Rees2016}, Genomics of Drug Sensitivity in Cancer (GDSCv1 and GDSCv2) \cite{Yang2012, Iorio2016, Picco2019},  Genentech Cell Line Screening Initiative (gCSI) \cite{Klijn2015, Haverty2016, Feizi2021}, National Cancer Institute (NCI-60) \cite{Shoemaker2006}, Profiling Relative Inhibition Simultaneously in Mixtures (PRISM) \cite{Corsello2020}, Finnish Institute for Molecular Medicine (FIMM) \cite{Mpindi2016}, as well as additional smaller studies of patient-derived organoids (PDOs) \cite{VANDEWETERING2015933, Lee2018BladderOrganoid, Narasimhan2020, Zhu2022Organoid}. For each study, we extracted raw drug response measurements, compound annotations, and associated metadata using the PharmacoGx R package \cite{Smirnov2015, RCoreTeam2021}.

A summary of all included datasets, along with their respective numbers of cancers, compounds, and drug–response experiments, is provided in Table~\ref{tab:data_sources}.

\begin{table}[ht]
\centering
\caption{Summary of pharmacogenomic datasets included in this study.}
\label{tab:data_sources}
\begin{tabular}{lcccc}
\hline
\textbf{Study} & \textbf{Cancer Type$^\ast$} & \textbf{\# Cancers} & \textbf{\# Drugs} & \textbf{\# Experiments} \\
\hline
CCLE\_2015        & Cell Line & 499 & 23    & 11,000 \\
CTRPv2\_2015      & Cell Line & 845 & 299   & 214,276 \\
FIMM\_2016        & Cell Line & 50  & 51    & 2,511 \\
GDSC\_2020 (v1)   & Cell Line & 986 & 249   & 267,731 \\
GDSC\_2020 (v2)   & Cell Line & 808 & 157   & 182,560 \\
gCSI\_2019        & Cell Line & 568 & 39    & 14,864 \\
NCI60\_2021       & Cell Line & 85  & 53,323 & 4,267,356 \\
PRISM\_2020       & Cell Line & 480 & 948   & 491,502 \\
Lee et al.\cite{Lee2018BladderOrganoid}               & PDO       & 11  & 47    & 1,329 \\
Narasimhan et al.\cite{Narasimhan2020}        & PDO       & 13  & 34    & 418 \\
Zhu et al. \cite{Zhu2022Organoid}            & PDO       & 47  & 34    & 1,195 \\
Van de Wetering et al.\cite{VANDEWETERING2015933}   & PDO       & 9   & 78    & 702 \\
Totals       & Both & 1,362 & 53,949   & 5,455,444 \\
\hline
\end{tabular}

\vspace{2mm}
\begin{flushleft}
\footnotesize{$^\ast$ PDO: Patient-derived organoid.}
\end{flushleft}
\end{table}

\subsection{2.2 Drug Response Data}

Drug response measurements were obtained from PharmacoDB as raw experimental readouts describing cell viability across multiple drug concentrations. These data originate from heterogeneous high-throughput screening studies that differ in assay protocols, concentration ranges, and reporting conventions. To enable consistent downstream modeling, all response data were processed using a unified workflow that prioritizes preservation of raw experimental information while enforcing standardized preprocessing and response quantification.

\subsubsection{2.2.1 Preprocessing and Quality Control}
\vspace{0.5em}
Exploratory analyses of response value distributions revealed the presence of extreme viability measurements in a subset of datasets, with occasional values far exceeding biologically plausible ranges. Rather than discarding affected experiments, response values were truncated to an upper bound corresponding to 200 percent viability prior to dose response curve fitting. This approach enabled retention of the full set of experiments while improving numerical stability during downstream modeling.

\subsubsection{2.2.2 Dose Response Modeling and Summary Metrics}
\vspace{0.5em}
Drug response was summarized by fitting a four-parameter logistic (4PL) dose--response model to normalized viability measurements as a function of log-transformed drug concentration. The model is defined as:

\begin{equation}
f(x) = E_{\infty} + \frac{E_0 - E_{\infty}}{1 + 10^{(x - EC_{50}) \cdot HS}}
\end{equation}

where $x$ denotes the log$_{10}$-transformed molar concentration, $E_0$ and $E_{\infty}$ represent the asymptotic response at zero and infinite dose, respectively, $EC_{50}$ is the concentration corresponding to the half-maximal response, and $HS$ is the hill slope governing curve steepness. Initial parameter bounds were selected to reflect biologically plausible drug response behavior while maintaining numerical stability during optimization. In particular, $E_0$ was effectively set at 1 to reflect normalized baseline viability, $E_{\infty}$ was bounded between 0 and 1, $EC_{50}$ was constrained within a broad concentration range, and the hill slope was allowed to vary to accommodate differences in response steepness across experiments. Under this initial parameterization, the bounds used were:
\[
E_{\infty}\in[0,1],\qquad
EC_{50}\in[-19,5],\qquad
HS\in[-5,5],\qquad
E_0=1
\]

Although most drug responses exhibit decreasing viability with increasing dose, a subset of experiments displayed increasing viability as a function of dose. When the resulting fit produced a negative hill slope, indicating an inverted dose--response relationship, the model was conditionally refit using the following relaxed bounds:
\[
E_{\infty}\in[0,\infty),\qquad
EC_{50}\in[-19,5],\qquad
HS\in[0,5],\qquad
E_0\in[0,\infty)
\]

which enforces a positive Hill slope while allowing both asymptotic response parameters to vary freely. Relaxing the asymptotic bounds removes the requirement that the response at zero dose exceed the response at infinite dose, allowing the model to represent increasing dose--response relationships. This preserves the functional form of the model while permitting increasing response curves when supported by the data. Selection between the original and inverted fits was based on goodness-of-fit, quantified by the coefficient of determination ($R^2$). The inverted fit was retained only if it improved $R^2$ relative to the original fit, ensuring that curve inversion was applied conservatively and only when justified by the observed data.

To summarize, two complementary fitting regimes were employed. In the \emph{strict} regime, parameter bounds enforced a canonical decreasing response and curve inversion was disabled. In the \emph{flexible} regime, parameter bounds were relaxed and curve inversion was enabled as described above. Area under the dose response curve (AUC) was computed under both regimes. In cases where flexible fitting produced extreme AUC values exceeding the theoretical maximum, AUC values were capped at 1.0.

\subsection{2.3 Chemical Data}

Chemical representations of drugs were standardized and encoded into multiple feature modalities to support downstream modeling. These included canonical SMILES strings, physicochemical descriptors, and binary molecular fingerprints. 

\subsubsection{2.3.1 Compound Representation and Curation}
\vspace{0.5em}
Raw SMILES strings were collected for all drugs across PharmacoDB datasets. To ensure consistency and chemical validity, SMILES strings were first cleaned to remove malformed entries, placeholder values, and formatting artifacts. Molecules were then parsed using RDKit \cite{RDKit}, and salts and counterions were removed using its SaltRemover function. Canonical SMILES were generated for all successfully parsed molecules to ensure a unique and standardized representation. In addition, tautomeric variants were identified and collapsed into canonical tautomers using RDKit. 

\subsubsection{2.3.2 Molecular Featurization}
\vspace{0.5em}
Two complementary molecular feature representations were derived from canonical SMILES strings. First, physicochemical descriptors were computed using the Mordred \cite{Moriwaki2018} descriptor library with three-dimensional descriptors disabled. This resulted in a high-dimensional feature vector capturing topological, constitutional, and electronic properties of each compound. Descriptor values that were undefined or failed to compute were recorded and subsequently zero-filled to maintain a consistent feature dimensionality across compounds. Second, structural fingerprints were generated using extended-connectivity fingerprints (ECFP4) with a radius of 2 and a fixed length of 512 bits. These binary fingerprints encode local atomic environments and are widely used for modeling structure--activity relationships.

\subsection{2.4 Multiomics Data}

The expanded resource contains six molecular profiling modalities that describe complementary aspects of cancer cell state: gene expression, copy number variation, somatic mutation, DNA methylation, protein expression (measured by the reverse phase protein array (RPPA)), and miRNA expression. Table~\ref{tab:multiomics_summary} summarizes the number of cancer samples and molecular features available for each modality, together with the number of samples that could be linked to at least one drug response measurement. The multiomics data were sourced from the Dependency Map (DepMap) portal of CCLE version 22Q2 (https://depmap.org/portal) \cite{Tsherniak2017}. Gene expression profiles were supplemented with data from PharmacoDB for cell lines not represented in the DepMap release. All molecular profiling data were processed using a similar methodology as the data included in the original IMPROVE benchmark dataset, providing expanded coverage while maintaining consistency with the original dataset \cite{Partin2026}.

Gene identifiers in the raw data were inconsistently reported across datasets, using a mixture of Ensembl gene identifiers (with version numbers), gene symbols, or Entrez IDs. To ensure compatibility with downstream models requiring specific identifier schemes, all genes were mapped using a curated gene information table derived from the National Center for Biotechnology Information (NCBI) \cite{OLeary2024}. Ensembl version suffixes were removed prior to mapping, and only genes with valid mappings across Ensembl, Entrez, and gene symbol identifiers were retained. Additionally, to enable integration across datasets and facilitate linkage with external annotations, cell line identifiers were mapped to Cellosaurus \cite{Bairoch2018} accession IDs using curated metadata files. This mapping enabled consolidation of synonymous cell line names and supported the creation of a unified sample information table containing cancer type annotations, aliases, and other relevant metadata. All omics and response data were indexed using these standardized identifiers.

\begin{table}[ht]
\centering
\caption{Summary of multi-omics datasets and overlap with drug response data.}
\label{tab:multiomics_summary}
\begin{tabular}{llrrr}
\hline
\textbf{Omics Type} & \textbf{Feature} & \textbf{\# Cancers} & \textbf{\# Features} & \textbf{\# in Response} \\
\hline
Copy Number              & Gene    & 1,764 & 25,331 & 1,261 \\
Gene Expression          & Gene    & 1,907 & 17,338 & 1,354 \\
Mutation Count           & Gene    & 1,769 & 18,739 & 1,257 \\
RPPA                     & Protein & 894   & 214    & 859   \\
miRNA Expression         & miRNA   & 946   & 734    & 893   \\
DNA Methylation          & TSS    & 839   & 19,606 & 803   \\
\hline
\end{tabular}
\begin{flushleft}
\footnotesize{$^\ast$ TSS: transcription start site.}
\end{flushleft}
\end{table}

\subsubsection{2.4.1 Gene Expression}
\vspace{0.5em}

Gene expression profiling measures the abundance of RNA transcripts and provides a representation of the transcriptional state of a cancer sample. Gene expression profiles were primarily obtained from DepMap. To increase coverage, RNA sequencing data for cell lines not represented in DepMap were extracted from PharmacoDB using PharmacoGx when transcript quantification was available. Raw molecular profiles from PharmacoDB were initially keyed by sequencing run identifiers and included technical and biological replicates. To obtain a single expression profile per cell line, replicate samples corresponding to the same cell line were collapsed by computing the mean expression value for each gene. Expression values were then converted to log$_2$(TPM + 1), where TPM refers to transcripts per million reads mapped, matching the representation used by DepMap. The processed PharmacoDB expression matrix was subsequently merged with the DepMap gene expression data. For cell lines present in both resources, DepMap profiles were retained in preference to the corresponding PharmacoDB profiles. Finally, quantile normalization was applied across the combined data to reduce distributional differences between data sources while preserving gene-wise rank information within each sample, enabling joint analysis of samples originating from distinct datasets.

\subsubsection{2.4.2 Copy Number Variation}
\vspace{0.5em}

Copy number variation (CNV) describes gains and losses of DNA relative to the normal diploid state. Gene-level copy-number data were curated from the DepMap 22Q2 data files \cite{Tsherniak2017}. These values were inferred from whole-genome sequencing, whole-exome sequencing, or SNP-array data, depending on availability \cite{Tsherniak2017}. Segment-level calls were mapped to genes, and a genomic-coordinate-weighted average was calculated for each gene \cite{Tsherniak2017}. The distributed values were represented as $\log_2(\mathrm{CN\ ratio}+1)$, where the CN ratio is the relative copy-number ratio of a genomic region compared with the genome-wide baseline. Discretized versions were also generated using five states: deep deletion ($-2$) for values $\leq 0.5210507$, heterozygous loss ($-1$) for values in $(0.5210507,0.7311832]$, diploid ($0$) for values in $(0.7311832,1.214125]$, gain ($1$) for values in $(1.214125,1.422233]$, and amplification ($2$) for values $>1.422233$.

\subsubsection{2.4.3 Mutation}
\vspace{0.5em}

Somatic mutation profiling identifies DNA sequence variants acquired during tumor development, including single-nucleotide variants (SNVs) and small insertions or deletions (indels). Depending on the application, mutation data may be represented either as aggregated gene-level features or as individual variant records containing genomic annotations. The gene-level count matrix records the number of mutation events observed in each gene for each cancer sample and is the representation summarized in Table~\ref{tab:multiomics_summary}. In addition, a long-format mutation table is included that preserves individual mutation events, with each record describing a single genomic variant together with associated annotations such as the affected gene, genomic coordinates, reference genome build, chromosome location, and mutation type. This data file retains mutation-level information for analyses requiring individual mutation records while complementing the aggregated gene-level mutation count matrix. 

\subsubsection{2.4.4 DNA Methylation}
\vspace{0.5em}

DNA methylation profiling measures the methylation state of cytosine residues within CpG dinucleotides, providing a representation of epigenetic regulation. The DepMap methylation data were generated using reduced-representation bisulfite sequencing (RRBS), a sequencing-based assay that enriches CpG-dense regions of the genome by digesting DNA with restriction enzymes followed by bisulfite conversion and sequencing. The resulting feature matrix contains methylation measurements for gene transcription start sites (TSS), with each feature corresponding to a TSS of a gene.

\subsubsection{2.4.5 Proteomics (RPPA)}
\vspace{0.5em}

Reverse phase protein array (RPPA) is an antibody-based proteomic assay that quantifies the abundance of selected proteins across many biological samples simultaneously. Each feature corresponds to a specific antibody target, allowing measurement of total protein levels or site-specific phosphorylation associated with cellular signaling pathways. Unlike transcriptomic data, RPPA directly measures protein abundance and post-translational modification for a targeted panel of proteins, providing a complementary molecular representation of cancer cell state.

\subsubsection{2.4.6 miRNA Expression}
\vspace{0.5em}

MicroRNAs (miRNAs) are short non-coding RNA molecules that regulate gene expression post-transcriptionally through interactions with messenger RNA. miRNA expression profiling quantifies the abundance of individual miRNAs within a sample, providing a representation of regulatory RNA activity that complements measurements of messenger RNA expression. As an additional molecular modality, miRNA expression can be incorporated into multi-omics studies to help capture regulatory processes not directly reflected by gene expression alone.

\subsection{2.5 Models}

We evaluated the impact of the expanded dataset using two DRP models: UNO \cite{Xia2022} and GraphDRP \cite{Nguyen2021}. The model architectures of UNO and GraphDRP are different. For example, UNO leverages predefined molecular descriptors of drugs, while GraphDRP learns from canonicalized SMILES via graph-based neural networks. 

\subsubsection{2.5.1 UNO}
\vspace{0.5em}
UNO is a deep neural network architecture that employs separate fully connected subnetworks to encode cell line and drug features independently. Drug features are represented using physicochemical descriptors computed with the Mordred library, providing a high-dimensional vector of interpretable molecular properties. These descriptor-based representations are processed through the drug encoder network, while gene expression profiles are processed through a separate cell line encoder. The resulting latent representations are then concatenated and passed through fully connected layers to predict drug response. Residual connections between neighboring layers are incorporated throughout the subnetworks to facilitate optimization and improve gradient propagation during training. 

\subsubsection{2.5.2 GraphDRP}
\vspace{0.5em}
GraphDRP is a graph-based deep learning model that represents drugs as molecular graphs, where atoms correspond to nodes and chemical bonds to edges. Graph convolutional networks are used to learn feature representations directly from these structures, enabling the model to capture local and global chemical relationships. Cell line features are encoded as fixed-length vectors, and then the learned representations of drugs and cell lines are concatenated and passed through a fully connected network to predict drug response. Gene expression was used as the input for cell line features for this analysis, although GraphDRP can also work with mutation and copy number variation data.

\subsection{2.6 Model Training and Evaluation}

To assess the impact of dataset expansion on DRP performance, each model was trained independently on both the original IMPROVE benchmark dataset and the expanded dataset constructed in this study. All models were evaluated using a shared test framework to enable a direct and controlled comparison between datasets.

\subsubsection{2.6.1 Data Splitting and Cross-Validation}
\vspace{0.5em}

Model performance was evaluated using 10-fold cross-validation with evaluation folds defined exclusively using experiments from three constituent data sources: GDSC, CCLE, and CTRPv2. These sources were selected because they are present in both the original and expanded benchmarks, allowing the effect of the additional training data to be assessed using identical test sets. Although the expanded benchmark additionally includes NCI60, PRISM, and FIMM, these sources were used only for training and were not used to define the test folds. To construct the folds, the unique drug identifiers and unique cell line identifiers present across GDSC, CCLE, and CTRPv2 were each randomly partitioned into 10 folds. For each cross-validation fold, all drug response experiments from GDSC, CCLE, and CTRPv2 involving the drugs assigned to the held-out drug fold formed the drug-blind test set, while all experiments involving the cell lines assigned to the held-out cell line fold formed the cancer-blind test set. Experiments in which both the drug and cell line belonged to the held-out folds formed a separate disjoint test set. The disjoint setting arises naturally from the independent holdout of drugs and cell lines and represents the most stringent generalization scenario. To ensure strict blind evaluation, all experiments involving the held-out drugs or cell lines were removed from the remaining data for that fold, regardless of their source. Consequently, experiments from GDSC, gCSI, CCLE, CTRPv2 and, for the expanded benchmark, NCI60, PRISM, and FIMM were excluded whenever they involved a held-out drug or cell line. This ensured that the test set remained identical for both benchmarks while preventing information leakage during training. After these exclusions, the remaining eligible experiments were randomly partitioned to generate a conventional mixed test set. Specifically, 10\% of the remaining experiments were assigned to the mixed test set, and an additional 10\% were randomly selected as a validation set for early stopping of model training. 

This resulted in the following test subsets:
\vspace{0.5em}
\begin{itemize}
    \item \textbf{Drug-blind:} experiments involving drugs not observed during training
    \vspace{0.5em}
    \item \textbf{Cancer-blind:} experiments involving cell lines not observed during training
    \vspace{0.5em}
    \item \textbf{Disjoint:} experiments in which both the drug and cell line were unseen during training
    \item \textbf{Mixed:} experiments where the drug and cell line were seen during training, but their pairing was not 
\end{itemize}
The training set consisted of all remaining eligible experiments after removal of the evaluation and validation sets. For the original benchmark, this comprised experiments from CCLE, GDSC, gCSI, and CTRPv2. For the expanded benchmark, the training set additionally included eligible experiments from NCI60, PRISM, and FIMM.

\subsubsection{2.6.2 Training Procedure}
\vspace{0.5em}

Both UNO and GraphDRP were trained using identical data splits and evaluation protocols. Training was performed on an 8-GPU computing cluster at Argonne National Laboratory. Default hyperparameters from prior implementations of each model were used to ensure consistency with established benchmarks \cite{Partin2026}. The batch size was increased to 512 to accommodate the larger scale of the expanded dataset. Models were optimized using mean squared error (MSE) loss. Early stopping was applied based on validation performance with a patience of 20 epochs, and training proceeded until convergence.

\section{3  Results} 
\begin{figure}[ht]
\centering

\begin{subfigure}[t]{0.48\textwidth}
    \centering
    \includegraphics[width=\textwidth]{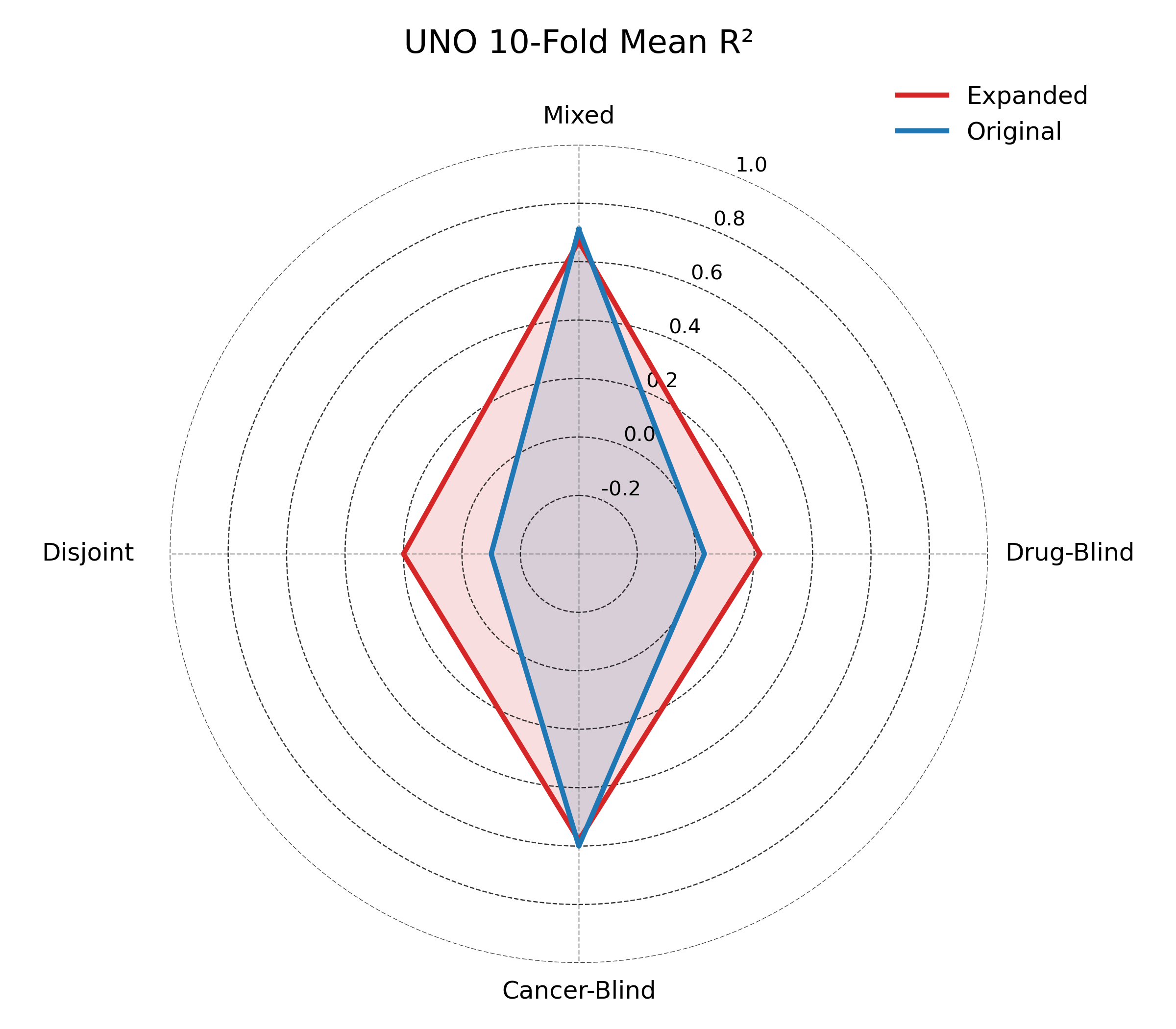}
    \caption{UNO}
\end{subfigure}
\hfill
\begin{subfigure}[t]{0.48\textwidth}
    \centering
    \includegraphics[width=\textwidth]{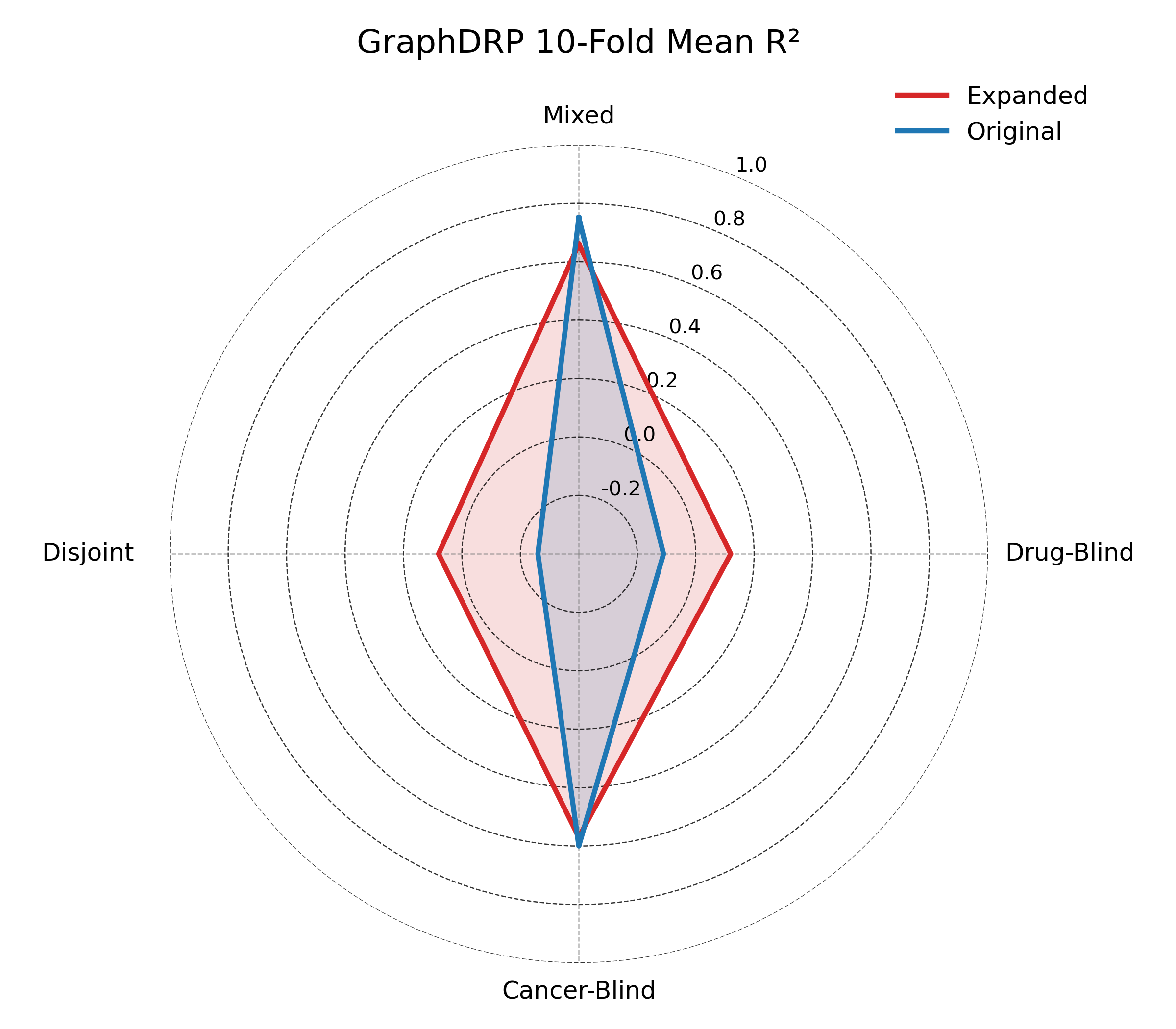}
    \caption{GraphDRP}
\end{subfigure}

\caption{Comparison of model performance across evaluation settings for models trained on the original and expanded datasets. Each subplot shows mean $R^2$ across four evaluation settings (mixed, drug-blind, cancer-blind, and disjoint).}
\label{fig:spider}
\end{figure}

Model performance was evaluated using the mean $R^2$ and Pearson correlation coefficient (PCC) across 10 cross-validation folds under four evaluation settings: mixed, cancer-blind, drug-blind, and disjoint. The average prediction performance of models across cross-validation trials are summarized in Tables~\ref{tab:results_r2_mean} and~\ref{tab:results_pcc_mean}, with the standard deviations summarized in Tables ~\ref{tab:results_r2_sd} and ~\ref{tab:results_pcc_sd}, respectively. Both models exhibited substantial improvements in the drug-blind and disjoint settings when trained on the expanded dataset. In contrast, performance slightly decreased in the mixed and cancer-blind settings. A potential reason for the slight performance drop in the mixed and cancer-blind cross-validation settings is that the test samples were drawn from the GDSCv2, CCLE, and CTRPv2 datasets, which were generated using experimental assays different from that used for the NCI60 dataset, while the NCI60 dataset constitutes the majority of the expanded dataset. Figure~\ref{fig:spider} provides a visual comparison of model performance across evaluation settings.

\begin{table}[ht]
\centering
\caption{Mean R² across 10 cross-validation folds.}
\label{tab:results_r2_mean}

\begin{tabular}{llcccc}
\hline
\textbf{Dataset} & \textbf{Model} & \textbf{Mixed} & \textbf{Drug-blind} & \textbf{Cancer-blind} & \textbf{Disjoint} \\
\hline
Original
& UNO
& 0.71
& 0.03
& 0.60
& -0.10 \\
& GraphDRP
& 0.75
& -0.11
& 0.60
& -0.26 \\
\hline
Expanded
& UNO
& 0.67
& 0.22
& 0.58
& 0.20 \\
& GraphDRP
& 0.66
& 0.12
& 0.57
& 0.08 \\
\hline
\end{tabular}
\end{table}

\begin{table}[ht]
\centering
\caption{Standard deviation of R² across 10 cross-validation folds.}
\label{tab:results_r2_sd}

\begin{tabular}{llcccc}
\hline
\textbf{Dataset} & \textbf{Model} & \textbf{Mixed} & \textbf{Drug-blind} & \textbf{Cancer-blind} & \textbf{Disjoint} \\
\hline
Original
& UNO
& 0.01
& 0.31
& 0.03
& 0.39 \\
& GraphDRP
& 0.01
& 0.19
& 0.02
& 0.27 \\
\hline
Expanded
& UNO
& 0.01
& 0.18
& 0.02
& 0.18 \\
& GraphDRP
& 0.01
& 0.28
& 0.02
& 0.28 \\
\hline
\end{tabular}
\end{table}

\begin{table}[ht]
\centering
\caption{Mean PCC across 10 cross-validation folds.}
\label{tab:results_pcc_mean}

\begin{tabular}{llcccc}
\hline
\textbf{Dataset} & \textbf{Model} & \textbf{Mixed} & \textbf{Drug-blind} & \textbf{Cancer-blind} & \textbf{Disjoint} \\
\hline
Original
& UNO
& 0.87
& 0.39
& 0.81
& 0.34 \\
& GraphDRP
& 0.87
& 0.21
& 0.79
& 0.14 \\
\hline
Expanded
& UNO
& 0.83
& 0.51
& 0.77
& 0.49 \\
& GraphDRP
& 0.82
& 0.45
& 0.76
& 0.42 \\
\hline
\end{tabular}
\end{table}

\begin{table}[ht]
\centering
\caption{Standard deviation of PCC across 10 cross-validation folds.}
\label{tab:results_pcc_sd}

\begin{tabular}{llcccc}
\hline
\textbf{Dataset} & \textbf{Model} & \textbf{Mixed} & \textbf{Drug-blind} & \textbf{Cancer-blind} & \textbf{Disjoint} \\
\hline
Original
& UNO
& 0.01
& 0.15
& 0.01
& 0.17 \\
& GraphDRP
& 0.01
& 0.16
& 0.01
& 0.16 \\
\hline
Expanded
& UNO
& 0.01
& 0.14
& 0.01
& 0.14 \\
& GraphDRP
& 0.01
& 0.16
& 0.01
& 0.17 \\
\hline
\end{tabular}
\end{table}

The trend in prediction performance was generally consistent across cross-validation trials, with improvements in the drug-blind and disjoint settings observed in the majority of trials for both models. While a minority of trials showed lower performance when trained on the expanded dataset, the overall trend remained positive, with average $R^2$ improvements clearly favoring the expanded dataset. Notably, one fold exhibited substantial variability, with markedly reduced performance when using the expanded dataset for GraphDRP, and the reverse trend for UNO. This behavior suggests sensitivity to specific data partitions in rare cases, but the overall pattern still shows improved generalization across data folds on average. Taken together, these results indicate that the observed gains are not driven by a small number of favorable data splits, but instead reflect a robust, though not completely uniform, improvement associated with increased dataset scale and chemical diversity.

\section{4 Discussion}

In this work, we present a substantial expansion of the standardized pharmacogenomic benchmark underlying the IMPROVE framework, with the goal of improving the generalization capacity of drug response prediction models. By integrating large-scale data from PharmacoDB and additional sources into a unified, model-ready format, we significantly increased both the volume of drug–response measurements and the diversity of chemical space represented in the dataset. Our results demonstrate that this expansion leads to consistent improvements in the most challenging DRP settings—drug-blind and disjoint—highlighting the importance of dataset scale and diversity for learning generalizable representations in DRP tasks.

The observed gains in drug-blind and disjoint settings suggest that increased chemical diversity enables models to better capture structure–activity relationships that extend beyond the compounds seen during training. This is particularly important in the context of drug discovery, where predictive models must extrapolate to novel chemical entities. The improvements across both UNO and GraphDRP indicate that this effect is not architecture-specific, but rather reflects a fundamental benefit of training on a broader and more representative dataset. At the same time, the relatively modest or slightly decreased performance in mixed and cancer-blind settings suggests that expanding dataset scale does not uniformly improve all aspects of predictive performance. One possible explanation is that the inclusion of more heterogeneous data introduces additional variability that may reduce performance on tasks involving familiar entities, even as it enhances generalization to unseen ones. 

An additional factor contributing to this behavior is the composition of the expanded dataset, which is heavily dominated by experiments from the NCI60 study. This introduces a distributional imbalance relative to the original benchmark, which was more evenly composed across datasets. As a result, models trained on the expanded dataset may become more biased toward patterns specific to NCI60, potentially reducing the test performance on other datasets. In particular, the shared test dataset—constructed from the CCLE, CTRPv2, and GDSC studies—was generated using a viability assay different from the one used in the NCI60 study, which possibly causes the reduced performance in the mixed set evaluation, even as the model generalizability in the chemical space improves. From this perspective, the observed trade-off between mixed and drug-blind or disjoint performance could be interpreted as a consequence of shifting the training distribution toward greater scale and chemical diversity, at the expense of balance across source datasets.

Despite these advantages, the expanded dataset introduces several limitations. Most notably, the scale of the data substantially increases the computational resources required for model training. Training on millions of drug–response measurements with high-dimensional molecular and transcriptomic features necessitates access to multi-GPU infrastructure and extended training times, which may limit accessibility. This computational burden also constrains the ability to perform extensive hyperparameter tuning or explore more complex model architectures, potentially limiting the extent to which the dataset’s full predictive potential can be realized. In addition, while the dataset has been carefully curated and standardized, it still aggregates data from heterogeneous experimental sources, which may introduce residual noise or systematic biases that are difficult to fully eliminate.

Another limitation is that, although additional omics modalities are present in the source studies, they were not fully leveraged in the present analysis. As a result, the current evaluation primarily reflects the contribution of transcriptomic and chemical features, leaving the potential benefits of multi-omics integration underexplored. Given the importance of genomic, epigenomic, and proteomic factors in modulating drug response, incorporating these data more directly into modeling frameworks represents an important opportunity for further improvement. Integrating complementary data types such as mutation profiles, copy number variation, and proteomics could enable richer representations of cancer cell states and potentially improve predictive performance, particularly in complex or heterogeneous settings. In addition, the scale of the dataset makes it well-suited for exploring pretraining strategies and foundation-model approaches, where models are first trained on large, diverse datasets and then fine-tuned for specific prediction tasks. Such approaches may help mitigate some of the trade-offs observed between generalization and in-distribution performance.

Overall, the expanded dataset presented here provides a stronger foundation for the development of DRP models that generalize to novel compounds, a critical requirement for real-world applications in drug discovery. By combining increased scale, standardized preprocessing, and diverse data sources, this resource advances the ability of the community to train and evaluate models in challenging generalization scenarios.

\section{Data and Code Availability}

\begin{enumerate}
    \item \textbf{Reproduce.} The code used to curate and process the dataset constructed in this paper is available here: \url{https://github.com/vincevelle/CSA-and-PharmacoGx-Integration} 
    
    \item \textbf{Data.} The full expanded dataset is available here: \url{https://figshare.com/articles/dataset/AI-Ready_Drug_Response_Dataset/31130641}
    
    \item \textbf{Models.} The models used in this study are available in these links: \url{https://github.com/JDACS4C-IMPROVE/GraphDRP/tree/develop}, \url{https://github.com/JDACS4C-IMPROVE/UNO/tree/develop}
    
    \item \textbf{Workflow.} The standardized AI workflow is available here: \url{https://github.com/JDACS4C-IMPROVE}
\end{enumerate}

\section{Acknowledgement}

This research was part of the Innovative Methodologies and New Data for Predictive Oncology Model Evaluation (IMPROVE) project under the National Cancer Institute - Department of Energy (NCI-DOE) collaboration program that has been funded in whole or in part with federal funds from the National Cancer Institute, National Institutes of Health, Task Order No. 75N91019F00134 and from the Department of Energy under Award Number ACO22002-001-00000.

\printbibliography

\end{document}